\documentclass[superscriptaddress,twocolumn,showpacs,amsmath,amssymb]{revtex4}

\usepackage{graphicx}

\renewcommand{\vec}[1]{\mbox{\boldmath $#1$}}

\begin{document}

\title{Pairing correlations in nuclei on the neutron-drip line}

\author{K. Hagino}
\affiliation{ 
Department of Physics, Tohoku University, Sendai, 980-8578,  Japan} 

\author{H. Sagawa}
\affiliation{
Center for Mathematical Sciences,  University of Aizu, 
Aizu-Wakamatsu, Fukushima 965-8560,  Japan}


\begin{abstract}
  Paring correlations in weakly bound nuclei on the edge of neutron drip line
is studied by using 
a three-body model. 
A density-dependent contact interaction is employed 
to calculate the ground state of 
halo nuclei 
$^{6}$He and $^{11}$Li, as well as a skin nucleus $^{24}$O. 
Dipole excitations in these nuclei are also studied within the same 
model. 
We point out that the di-neutron type correlation 
plays a dominant role in the halo nuclei $^{6}$He and $^{11}$Li
having the coupled spin of the two neutrons 
$S$=0, while the 
correlation similar
to the BCS type is important in $^{24}$O. Contributions of
the spin $S$=1 and S=0 configurations
are separately discussed
in the low energy dipole excitations. 
\end{abstract}


\maketitle

\section{Introduction}
  It is now feasible to study the structure of nuclei on the edge of
  neutron drip line. Such nuclei are expected to have
  unique properties influenced by the large spatial distribution 
of weakly bound valence neutrons, 
for examples, halo, skin, new magic numbers and strong soft dipole 
excitations.

  Two-neutron halo nuclei (sometimes referred to as borromean nuclei
  when there is no bound state between a valence neutron and a core
  nucleus) like $^{6}$He and $^{11}$Li have been 
often  described as three-body systems 
consisting of two valence neutrons interacting with each other, and 
with the core \cite{BF1,BF2,BF3,VMP96,BVM97,Zhukov93,Fed95}. 
The three-body Hamiltonian with realistic two-body 
interactions has been solved by the Faddeev method \cite{Zhukov93,Fed95}. 
On the other hand, Bertsch and Esbensen have developed a three-body model with 
a density-dependent delta interaction among the valence 
neutrons \cite{BF1}. They have subsequently extended their model by 
taking into account 
the effect of the recoil of the core nucleus\cite{BF3}.
They showed that the density-dependent contact force 
well reproduces the results of Faddeev calculations even though the radial 
dependence of the adopted interactions are quite different \cite{BF3}.
To the date, the most sophisticated many-body calculations of light 
nuclei include also three-body forces which play an important 
role in obtaining the correct binding energies of light nuclei \cite{Pud95}. 
To a large extent, 
such three-body forces can also be simulated effectively by a 
density dependent force.

Recently, a Hartree-Fock-Bogoliubov (HFB) model 
has been applied to study a di-neutron structure of the drip line nuclei. 
The dipole response has also been studied with the quasi-particle 
random phase approximation (QRPA) taking into account the continuum effect
\cite{Matsuo}.  Cluster models have also been often used to study the 
borromean nuclei including the dipole excitations \cite{Aoyama,Myo}.

In this paper, we undertake a detailed discussion on the 
ground state as well as the dipole response 
of neutron-rich nuclei using a three-body model 
with the density-dependent delta force, paying special attention 
to the di-neutron structure of valence neutrons. 
We particularly study the 
borromean nuclei, $^{6}$He and $^{11}$Li, and also another 
drip line nucleus $^{24}$O as a comparison. 
Since $^{23}$O is bound, 
$^{24}$O is not a borromean nucleus by definition, 
although the root-mean-square
(rms) radius indicates a feature of very extended neutron wave functions. 
The model we use is essentially the same as that of
Bertsch and Esbensen\cite{BF1,BF2,BF3}, 
while  the interaction  is adjusted to fit the
separation energy of each drip line nucleus.

The paper is organized as follows.  In Sec. II, we discuss 
the three-body model and the adopted two-body 
interactions. 
The results for the borromean nuclei are compared with those for 
the drip line nucleus  $^{24}$O in Sec. III. 
A summary is given in Sec. IV.

\section{Three-body model}

We consider a three-body system consisting of two valence 
neutrons, and an inert core nucleus with  the mass number $A_c$. 
We use the same three-body Hamiltonian as in 
Ref. \cite{BF3}. That is, 
\begin{equation}
H=\hat{h}_{nC}(1)+\hat{h}_{nC}(2)+V_{nn}+\frac{\vec{p}_1\cdot\vec{p}_2}{A_cm}. 
\label{3bh}
\end{equation}
Here, $\hat{h}_{nC}$ is the single-particle Hamiltonian for a valence 
neutron interacting with the core and is given by 
\begin{equation}
\hat{h}_{nC}=\frac{p^2}{2\mu}+V_{nC}(r), 
\end{equation}
where $\mu=mA_c/(A_c+1)$ is the reduced mass. 
The reduced mass $\mu$, together with the last term in
Eq. (\ref{3bh}), originate from the recoil kinetic energy of the
core\cite{BF3}. 
$V_{nn}$ is the interaction between the valence neutrons given by 
\begin{equation}
V_{nn}(\vec{r}_1,\vec{r}_2)=\delta(\vec{r}_1-\vec{r}_2)
\left(v_0+\frac{v_\rho}{1+\exp[(r_1-R_\rho)/a_\rho]}\right). 
\label{vnn}
\end{equation}
It is well known that 
the delta force (\ref{vnn}) must be supplemented with an energy cutoff 
$E_{\rm cut}$ in the two-particle spectrum. 
In terms of the energy cutoff $E_{\rm cut}$ and the scattering length
$a_{nn}$ for
$nn$ scattering, the strength for the delta interaction $v_0$ is 
given by \cite{BF3}
\begin{equation}
v_0=\frac{2\pi^2\hbar^2}{m}\,\frac{2a_{nn}}{\pi-2k_c\,a_{nn}},
\end{equation}
where $E_{\rm cut}=\hbar^2k_c^2/m$. 
The parameters for the density dependent part, i.e., 
$v_\rho, R_\rho$, and $a_\rho$, are adjusted in order to 
reproduce the known ground state properties for each nucleus. 
We specify the value of the parameters below.

We diagonalize the  Hamiltonian (\ref{3bh})
in the model space of the two-particle states with the energy 
$\epsilon_1+\epsilon_2\leq (A_c+1)E_{\rm cut}/A_c$ \cite{BF3}, 
where $\epsilon$ is  a single-particle energy of valence particle.
We use a Woods-Saxon potential for $V_{nC}$ 
to generate the single-particle basis:
\begin{eqnarray}
V_{nC}(r)&=&V_0\left(1-0.44f_{so}r_0^2(\vec{l}\cdot\vec{s})
\frac{1}{r}\frac{d}{dr}\right)\nonumber \\
&&~~~\times \left[1+\exp\left(\frac{r-R}{a}\right)\right]^{-1},
\end{eqnarray}
where $R=r_0A_c^{1/3}$. 
For $^6$He, we use the parameter set 
$a$=0.65 fm, $r_0$=1.25 fm, $V_0=-47.4$ MeV, and $f_{so}$=0.93, 
that reproduces the measured low-energy $n$-$\alpha$ phase shifts
\cite{BF3}. We employ the same parameters 
for the density-dependent interaction as in the line 5 of Table II in 
Ref. \cite{BF3}:
$a_{nn}=-15$ fm, 
$E_{\rm cut}=40$ MeV, $v_\rho=-v_0$, $R_\rho$=2.436 fm, and 
$a_\rho$=0.67 fm. The continuum single-particle spectrum is 
discretized with a radial box of $R_{\rm box}=30$ fm.

For  $^{11}$Li, 
we use 
$a$=0.67 fm, $r_0$=1.27 fm, $f_{so}$=1.006, and $R_{\rm box}=40$ fm. 
For the Woods-Saxon potential, 
a deep potential    $V_0=-47.5$ MeV is used for the even parity states, 
while a shallow  potential $V_0=-35.366$ MeV is adopted for the 
odd parity states 
in order to 
increase the $s$-wave component of the ground state wave function \cite{BF3}.
Similar potentials are used also  in Ref. \cite{VMP96}.
For the density dependent force, we use a similar parameter set 
$a_{nn}=-15$ fm, 
$E_{\rm cut}=30$ MeV, $v_\rho=-v_0$, $R_\rho$=2.935 fm, and 
$a_\rho$=0.67 fm to that of
the line 5 in Table IV in Ref. \cite{BF3} except the 
value of the energy cutoff.

For  $^{24}$O, 
we use $V_0=-43.2$ MeV, $a$=0.67 fm, $r_0$=1.25 fm, 
$f_{so}$=0.73, and $R_{\rm box}=30$ fm so that the bound s$_{1/2}$ and 
d$_{5/2}$ states have  empirical 
single-particle energies $-2.739$ and $-3.806$ MeV observed in $^{23}$O
and $^{21}$O, respectively. 
For the pairing interaction, we use 
$a_{nn}=-15$ fm, 
$E_{\rm cut}=30$ MeV, $v_\rho=814.2$ MeV fm$^3$, $R_\rho=R$, and 
$a_\rho$=0.67 fm which reproduce 
the two neutron separation energy of $^{24}$O, S$_{2n}$=6.452 MeV.

The calculated ground state properties are summarized in Table 1,
where 
\begin{equation}
\langle r_{nn}^2\rangle =\langle \Psi_{gs}|(\vec{r}_1-\vec{r}_2)^2
|\Psi_{gs}\rangle,
\end{equation}
is the mean square distance between the valence neutrons, and 
\begin{equation}
\langle r_{c-2n}^2\rangle =\langle \Psi_{gs}|(\vec{r}_1+\vec{r}_2)^2/4
|\Psi_{gs}\rangle,
\end{equation}
is the mean square distance of their center of mass with respect to
the core.

\begin{table}[hbt]
\caption{
Ground state properties of $^6$He, $^{11}$Li, and $^{24}$O 
obtained with the three-body model with the density-dependent delta 
interaction. The result for 
$^6$He is the same as 
the line 5 in Table II in Ref. \cite{BF3}.}
\begin{center}
\begin{tabular}{cccccc}
\hline
\hline
nucleus & $S_{2n}$ & $\langle r_{nn}^2 \rangle$ & 
$\langle r_{c-2n}^2 \rangle$ & dominant 
& fraction \\
& (MeV) & (fm$^2$) &
  (fm$^2$) & configuration 
&  (\%) \\
\hline
$^6$He & 0.975 & 21.3 & 13.2 & (p$_{3/2})^2$ & 83.0 \\
$^{11}$Li & 0.295 & 41.4 & 26.3 & (p$_{1/2})^2$ & 59.1 \\
$^{24}$O & 6.452 & 35.2 & 10.97 & (s$_{1/2})^2$ & 93.6 \\
\hline
\hline
\end{tabular}
\end{center}
\end{table}

\section{Discussions}

\subsection{Ground state properties}

Let us now discuss the spatial correlation of the valence neutrons in 
the ground state, and its influence to the dipole excitations near the 
neutron threshold. To this end, we first plot the two-particle density. It is 
given as a function of 
two radial coordinates, $r_1$ and $r_2$, for the valence neutrons, and the 
angle between them, $\theta_{12}$. 
The two-particle density can be decomposed into the $S=0$ and $S=1$ 
components in the $LS$-coupling scheme, i.e., 
\begin{equation}
\rho_2(r_1,r_2,\theta_{12})=\rho^{S=0}_2(r_1,r_2,\theta_{12})
+\rho^{S=1}_2(r_1,r_2,\theta_{12}).
\end{equation}
The explicit expression for the each component is given by \cite{BF1}
\begin{eqnarray}
&&\rho^{S=0}_2(r_1,r_2,\theta_{12}) =  \nonumber \\
&&
\frac{1}{8\pi}\sum_L
\sum_{l,j}\sum_{l',j'}\frac{\hat{l}\hat{l'}\hat{L}}{\sqrt{4\pi}}
\left(\begin{array}{ccc}
l&l'&L \\
0&0&0
\end{array}
\right)^2 \nonumber \\
&&~\times\Phi_{lj}(r_1,r_2)\Phi_{l'j'}(r_1,r_2)Y_{L0}(\theta_{12}) 
\nonumber \\
&&~\times(-)^{l+l'}\sqrt{\frac{2j+1}{2l+1}}\sqrt{\frac{2j'+1}{2l'+1}},
\label{rho2s0}
\\
&&\rho^{S=1}_2(r_1,r_2,\theta_{12})=  \nonumber \\
&&
\frac{1}{8\pi}\sum_L
\sum_{l,j}\sum_{l',j'}\frac{\hat{l}\hat{l'}\hat{L}}{\sqrt{4\pi}}
\left(\begin{array}{ccc}
l&l'&L \\
0&0&0
\end{array}
\right) 
\left(\begin{array}{ccc}
l&l'&L \\
1&-1&0
\end{array}
\right) \nonumber \\
&&~\times\Phi_{lj}(r_1,r_2)\Phi_{l'j'}(r_1,r_2)Y_{L0}(\theta_{12}) 
\nonumber \\
&&~\times(-)^{j+j'}\sqrt{2-\frac{2j+1}{2l+1}}\sqrt{2-\frac{2j'+1}{2l'+1}},
\label{rho2s1}
\end{eqnarray}
where $\hat{l}=\sqrt{2l+1}$. 
Here, 
$\Phi_{lj}(r,r')$ is the radial part of the two-particle wave function 
defined as
\begin{eqnarray}
\Phi_{lj}(r,r')&=&\sum_{n'\leq n}\frac{\alpha_{nn'lj}}
{\sqrt{2(1+\delta_{n,n'})}} \nonumber \\
&\times& (\phi_{nlj}(r)\phi_{n'lj}(r')+\phi_{nlj}(r')\phi_{n'lj}(r)),
\end{eqnarray}
where $n$ and $n'$ are the radial quantum numbers and 
$\alpha_{nn'lj}$ is the expansion coefficient. 
$\phi_{nlj}(r)$ is the radial part of the Woods-Saxon
   single particle wave function. 
Notice that the two-particle density is normalized as 
\begin{equation}
\int_0^\infty 4\pi r_1^2 \,dr_1 
\int_0^\infty r_2^2 \,dr_2 \int_0^\pi 2\pi \sin\theta_{12}\,d\theta_{12} \,
\rho_2(r_1,r_2,\theta_{12})=1.
\end{equation}

\begin{figure}
\includegraphics[scale=1.1,clip]{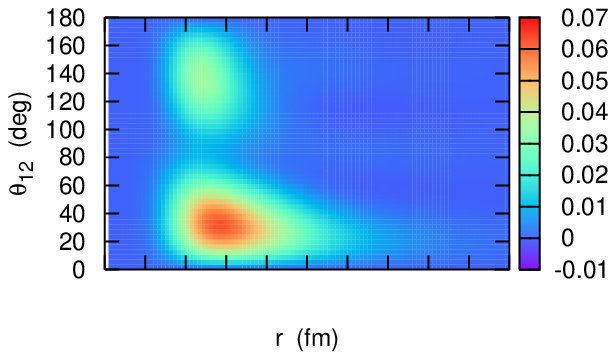}\\
\vspace{-1.5cm}
\includegraphics[scale=1.1,clip]{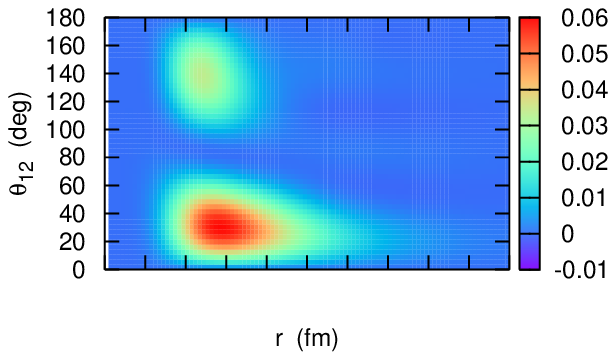}\\
\vspace{-1.5cm}
\includegraphics[scale=1.1,clip]{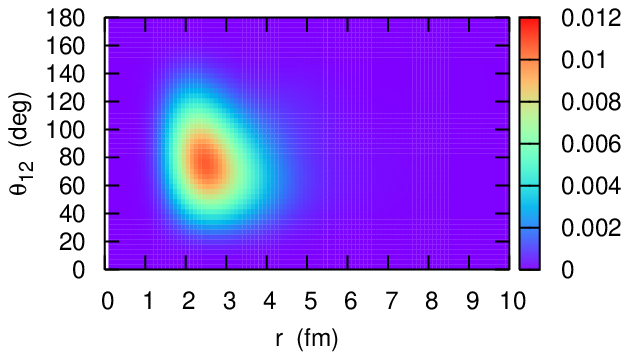}
\caption{(Color online) 
The two-particle density for $^6$He
  as a function of $r_1=r_2=r$ and the angle between 
the valence neutrons, $\theta_{12}$. 
The two-particle density is weighted with a factor 
$4\pi r^2\cdot 2\pi r^2\sin\theta_{12}$. 
The top panel shows 
the total density, while the middle and the bottom panels show 
the $S$=0 and the $S$=1 components in the $LS$-coupling scheme,
respectively. 
}
\end{figure}

\begin{figure}
\includegraphics[scale=1.1,clip]{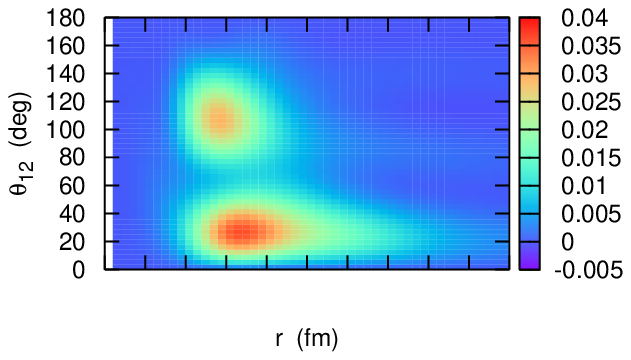}\\
\vspace{-1.5cm}
\includegraphics[scale=1.1,clip]{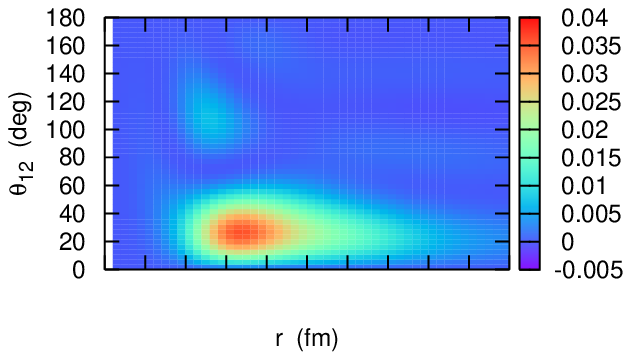}\\
\vspace{-1.5cm}
\includegraphics[scale=1.1,clip]{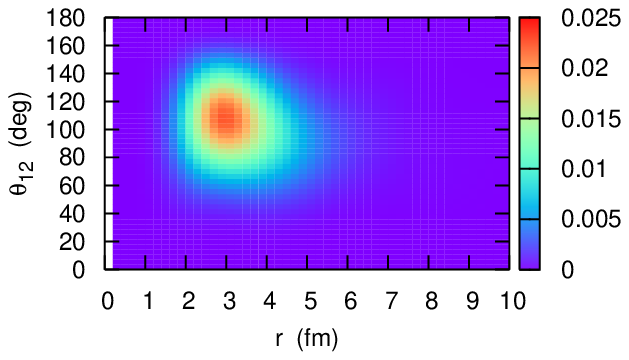}
\caption{(Color online) 
Same as fig. 1, but for $^{11}$Li. 
}
\end{figure}

\begin{figure}
\includegraphics[scale=1.1,clip]{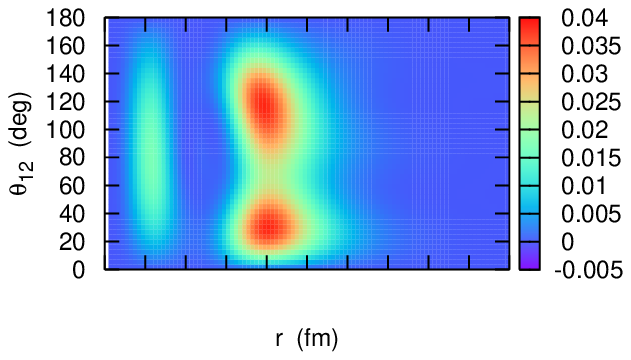}\\
\vspace{-1.5cm}
\includegraphics[scale=1.1,clip]{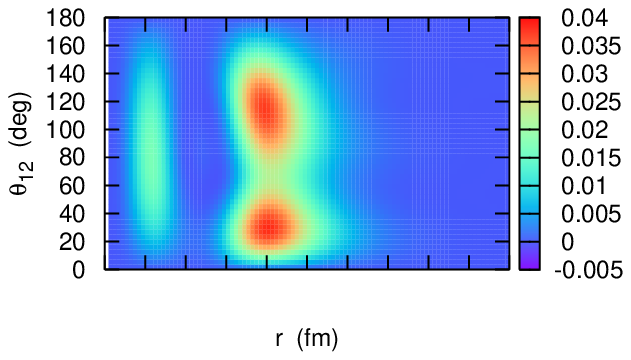}\\
\vspace{-1.5cm}
\includegraphics[scale=1.1,clip]{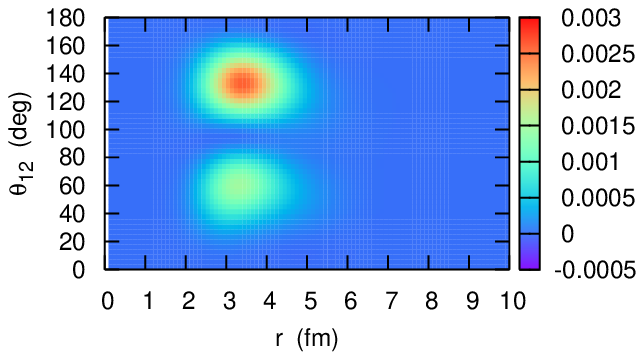}
\caption{(Color online) 
Same as fig. 1, but for $^{24}$O. 
}
\end{figure}

Figures 1,2, and 3 show the (total) two-particle density (the top panels) 
for the $^6$He, $^{11}$Li, and $^{24}$O nuclei, respectively, and their 
spin decompositions (the middle and the bottom panels). 
These are plotted as a function of the radius $r_1=r_2\equiv r$ and 
the angle $\theta_{12}$, and with a weight of 
$4\pi r^2\cdot 2\pi r^2 \sin \theta_{12}$. 
As has been pointed out in Refs. \cite{BF1,Zhukov93,OZV99}, one 
observes two peaks in the two-particle densities, 
although the two peaked structure is 
somewhat smeared in $^{24}$O. The peaks at smaller and larger 
$\theta_{12}$ are referred to as ``di-neutron'' and ``cigar-like'' 
configurations in Refs. \cite{Zhukov93,OZV99}, respectively. 
We see that the di-neutron part of the two-particle density 
has a long radial tail in 
$^{6}$He and $^{11}$Li, and thus can be interpreted as a halo structure. 
In contrast, the cigar-like configuration has a rather compact radial 
shape. 
For $^{24}$O, 
the di-neutron and the cigar-like configurations behave similarly as a 
function of $r$, and do not show a halo structure. 
Evidently, a large rms radius of $^{24}$O is attributed to 
the dominant $s$-wave component in the ground state wave function, 
rather than the halo effect (see below).

We find that 
the spin structure of the two-particle density 
is considerably different among the three nuclei studied. 
In order to see this transparently, 
we introduce the angular density $\rho(\theta_{12})$ 
by integrating the radial coordinates in the two-particle density, i.e., 
\begin{equation}
\rho(\theta_{12})\equiv 
4\pi\int_0^\infty r_1^2 \,dr_1 
\int_0^\infty r_2^2 \,dr_2 \rho_2(r_1,r_2,\theta_{12}). 
\label{rhoang}
\end{equation}
The angular density is normalized to unity as 
\begin{equation}
2\pi\int_0^\pi \sin\theta_{12}\,d\theta_{12} \,\rho(\theta_{12})=1. 
\end{equation}
Fig. 4 shows the angular density for 
the $^6$He, $^{11}$Li, and $^{24}$O (with a weight of $2\pi\sin\theta_{12}$). 
The solid line is the total density, while the dashed and the dotted lines 
are for the $S=0$ and the $S=1$ components, respectively. 
The expectation value of the angle $\theta_{12}$ is 
66.33, 65.29, and 82.37 degree. for 
$^6$He, $^{11}$Li, and $^{24}$O, respectively. 
For the borromean nuclei 
$^{6}$He and  $^{11}$Li, 
the $S$=0 wave function
dominates the di-neutron part of the two-particle density. 
In contrast, 
the cigar-like part has a large $S$=1 part in 
$^{11}$Li, but still the $S$=0 component dominates in $^{6}$He. 
For the $^{24}$O nucleus, 
there is no clear separation between the di-neutron 
and the cigar-like type structures. 
The wave function shows a strong correlation typical in the BCS 
type wave function. In fact, 
the calculated rms radius for $^{24}$O, 4.45 fm, is close 
to that of the bound 2s$_{1/2}$ state, 4.65 fm. 
A small difference
in the rms radii is 
due to the 
anti-halo effect, caused by the 
pairing correlation \cite{BDP00}.

\begin{figure}
\includegraphics[scale=0.4,clip]{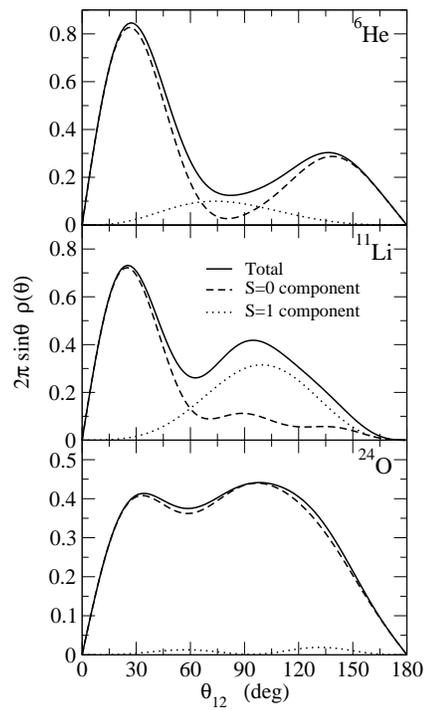}
\caption{
The angular density defined by Eq. (\ref{rhoang}) 
(weighted with a factor $2\pi\sin\theta$) 
for $^6$He, $^{11}$Li, and $^{24}$O. 
The solid line is for the total density, 
while the dashed and the dotted lines are for 
the $S$=0 and the $S$=1 components in the $LS$-coupling scheme,
respectively. 
}
\end{figure}

The main features of the angular dependence of the 
two-particle density shown in Fig. 4 
can be understood in the following way. 
From Eqs. (\ref{rho2s0}) and (\ref{rho2s1}), one can obtain by 
inserting the values of 3$j$-symbols  that 
\begin{eqnarray}
\rho^{S=0}(\theta_{12})&\propto& \frac{1}{3}Y_{00}(\theta_{12})+
\frac{2\sqrt{5}}{15}Y_{20}(\theta_{12}) \propto
\cos^2\theta_{12} \label{2nS0} \\
\rho^{S=1}(\theta_{12})&\propto& \frac{1}{3}Y_{00}(\theta_{12})-
  \frac{5}{\sqrt{15}}Y_{20}(\theta_{12}) \propto\sin^2\theta_{12}
  \label{2nS1}
\end{eqnarray}
for 
the configurations 
$(j,l)=(j',l')=$p$_{3/2}$ or 
$(j,l)=(j',l')=$p$_{1/2}$ . 
When weighted by $\sin\theta_{12}$, Eq. (\ref{2nS0}) has a peak at 
$\theta_{12}$=35.26 degree and 144.74 degree, while Eq. (\ref{2nS1}) 
has the maximum at 
$\theta_{12}$=90 degree. This is indeed the case for the 
$^6$He nucleus. For the $^{11}$Li, the admixture of
the (s$_{1/2})^2$ and (d$_{5/2})^2$ configurations
  perturb this picture, and 
a peak at $\theta_{12}$=144.74 degree
   in the $S$=0 component 
disappears to a large extent. 
For the $^{24}$O nucleus, the $S=1$ component is largely suppressed since 
the pure (s$_{1/2})^2$ state cannot form the $S=1$ configuration. 
Also, the two-particle density for the pure (s$_{1/2})^2$ configuration 
is proportional to $|Y_{00}|^2$, and thus has a peak at $\theta_{12}$=90 
degree when it is weighted by $\sin\theta_{12}$.

\subsection{Dipole excitations}

\begin{figure}
\includegraphics[scale=0.4,clip]{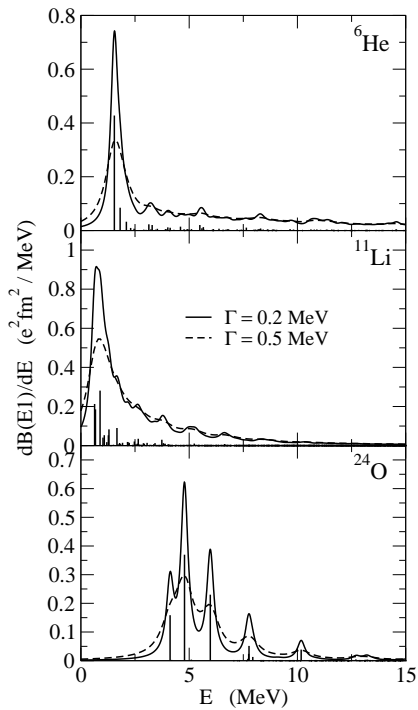}
\caption{
The B(E1) distribution for 
$^6$He, $^{11}$Li, and $^{24}$O. 
The solid and the dashed lines are obtained with a 
smearing procedure with $\Gamma=0.2$ and 0.5 MeV, respectively. 
}
\end{figure}

We next discuss the response of the ground state to the dipole 
field, 
\begin{equation}
\hat{D}_M=-\frac{Z}{A}e\,
(r_1Y_{1M}(\hat{\vec{r}}_1)+r_2Y_{1M}(\hat{\vec{r}}_2)).
\end{equation}
Since we obtain the excited 1$^-$ states by the matrix diagonalization, 
they appear as discrete states. We smear the discrete strength distribution 
with a smearing function as 
\begin{equation}
B(E1)=\sum_k\,\frac{\Gamma}{\pi}\,\frac{1}{(E-E_k)^2+\Gamma^2}\,B_k(E1),
\label{smear}
\end{equation}
where $B_k(E1)$ is the $B(E1)$ strength for the $k$-th excited state, 
\begin{equation}
B_k(E1)=3\,|\langle\Psi_{1^-}^k|\hat{D}_0|\Psi_{gs}\rangle|^2.
\end{equation}
Fig. 5 shows the $B(E1)$ distributions for 
the $^6$He, $^{11}$Li, and $^{24}$O. The solid and the dashed lines 
are obtained with the smearing function (\ref{smear}) with 
$\Gamma$=0.2 and 0.5 MeV, respectively. 
The discrete distributions are also shown. 
The total $B(E1)$ strength, $\sum_kB_k(E1)$, is 1.31, 1.76 and 0.97 
$e^2$ fm$^2$ for the $^6$He, $^{11}$Li, and $^{24}$O, respectively.

We notice that the strong threshold peak appears 
in the response of $^{6}$He and $^{11}$Li. 
In the cluster model within the plane wave approximation, the 
peak of the strength function appears at 1.6$S_c$, where $S_c$ is the 
cluster separation energy \cite{STG92,BBH91,HHB03}.
The calculated peaks in Fig. 5 are  at 1.55 and 0.66 MeV for 
$^6$He and $^{11}$Li, respectively. 
These peaks  are very close to 1.6 times 
the two neutron 
separation energy, 1.6$S_{2n}$=1.56 MeV for $^6$He and 0.47 MeV for $^{11}$Li. 
This similarity suggests the existence of 
strong di-neutron 
correlations in these nuclei. 
A small difference between the peak energy in Fig. 5 and 1.6$S_{2n}$ 
for $^{11}$Li is due to the large configuration mixing of s$_{1/2}$ state
 (22.7\%) in the ground state.
In contrast, for $^{24}$O, the peak (4.78 MeV) 
is below the two neutron 
separation energy, and is rather close to 1.6 times 
single particle energy for the 2s$_{1/2}$ state, that is, 1.6$\times$ 2.74 
= 4.38 MeV. Therefore, the di-neutron correlation does not seem to 
play a major 
role in the dipole response in this nucleus.

\begin{figure}
\includegraphics[scale=0.4,clip]{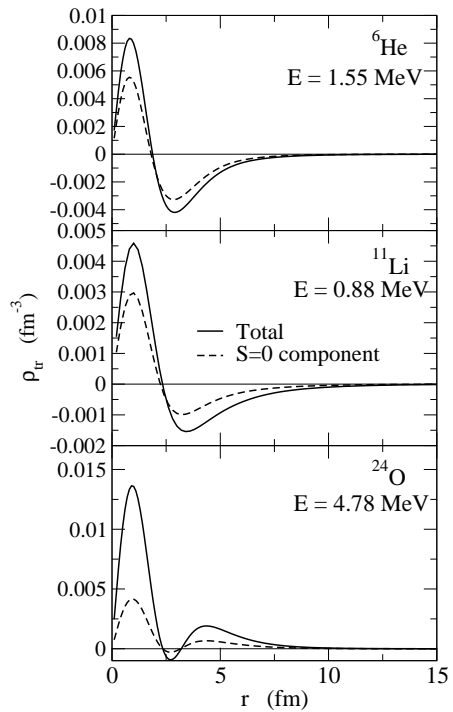}
\caption{
The transition density for the 
state close to the peak in the the B(E1) distribution. 
The top, middle, and bottom panels are 
for $^6$He, $^{11}$Li, and $^{24}$O, respectively. 
The solid line indicates the total density, while the 
$S$=0 component is denoted by the dashed line. 
}
\end{figure}

In order to see the threshold effect more clearly, we plot in Fig. 6 
the transition density for the strong E1 peaks for each nucleus. 
As a comparison, we also show the $S$=0 component of the transition 
density by the dashed line. 
For 
$^6$He and $^{11}$Li, the transition density shows a nodal structure and 
changes its sign, that is typical in the coupling to the continuum spectrum. 
Also, the $S$=0 component plays a significant role, supporting the 
importance of the di-neutron correlation in these nuclei. 
On the other hand, such clear nodal structure is not seen in the 
transition density of $^{24}$O. 
The transition density seems to consist of a coherent sum of the $S$=0 and 
the $S$=1 components. 
The dipole excitation in $^{24}$O is therefore interpreted as 
a coherent superposition of particle-hole excitations, 
rather than the continuum excitations, 
as in stable 
nuclei where the continuum effect is much less important.

We mention that, 
while the Coulomb breakup for the one-neutron halo $^{11}$Be is now 
well established, that for the two-neutron halo $^{11}$Li is still in dispute
  showing large discrepancies between the experimental data taken by
  three different groups \cite{Li11_exp}.
  Recently, Nakamura and his collaborators performed the Coulomb 
dissociation experiments of $^{11}$Li on $^{208}$Pb target with much higher 
statistics and with much less ambiguities caused by cross talk events in 
detecting two neutrons \cite{NF05}.
   They observed a sharp peak at E$_{\rm exp}\sim$0.6MeV and the integrated 
strength
   B$_{\rm exp}$(E1)=1.5$\pm 0.1 e^2$ fm$^2$ for E$ <$ 3.3MeV,  
which are consistent with
   the present results  E$_{\rm peak}$=0.66 MeV with the calculated strength
  B$_{\rm cal}$(E1)=1.31 $e^2$ fm$^2$ for E$ <$ 3.3MeV. 
Similar calculated results were also reported in Ref. \cite{BF3}.

\section{SUMMARY}

We have studied the role of di-neutron correlations in weakly bound
nuclei on the neutron drip line using a three-body model. 
The model Hamiltonian consists of a Woods-Saxon potential between a
valence neutron and the core, and of a density-dependent pairing 
interaction among the valence neutrons. We applied this model to 
the borromean nuclei, $^6$He and $^{11}$Li, as well as a skin nucleus 
$^{24}$O. 
For the borromean nuclei, $^6$He and $^{11}$Li, we found that the 
two-particle density has a two peaked structure, one peak at a small 
opening angle between the valence neutrons and the other at a large 
angle (the 'di-neutron' and 'cigar-like' configurations, 
respectively). We found that the former is dominated by the $S$=0
configurations in the $LS$ coupling scheme and has a long tail. 
On the other hand, the latter has a compact shape, and is dominated by 
the $S$=1 configuration for $^{11}$Li while 
by the $S$=0 configuration for $^{6}$He. 
For $^{24}$O, there is no clear separation between the 
di-neutron and the cigar-like configurations, and the ground state is 
dominated by the 
$S$=0 configuration.

We have also studied the dipole response of these nuclei within 
the same model. 
We found 
strong threshold peaks in the response of  $^{6}$He and  $^{11}$Li
nuclei, where the transition density shows the importance of the
coupling to the continuum.
The $S$=0 configuration, thus the di-neutron correlation, was found 
to have a large 
contribution to the transition density for these peaks. 
On the other hand, 
no clear sign of the continuum coupling was seen in the 
response of $^{24}$O. 
The transition density for the low energy dipole 
strength of $^{24}$O 
consists of a coherent sum of the $S$=0 and 
the $S$=1 components, and 
therefore the di-neutron correlation plays a much less important 
role in $^{24}$O than in the borromean nuclei.

Recently, a new Coulomb breakup measurement for $^{11}$Li has been 
undertaken at RIKEN \cite{NF05}. 
It would be interesting to perform a similar experiment also for 
the $^{24}$O nucleus and see a difference between them as we discussed 
in this paper.

\begin{acknowledgments}
This work was supported by the Japanese
Ministry of Education, Culture, Sports, Science and Technology
by Grant-in-Aid for Scientific Research under
the program numbers (C(2)) 16540259 and 16740139.
\end{acknowledgments}


\begin{thebibliography}{99}

\bibitem{BF1}
G.F.  Bertsch and  H. Esbensen, Ann. Phys. (N.Y.) {\bf 209}, 327
(1991).

\bibitem{BF2}  H. Esbensen  and  G.F. Bertsch, Nucl. Phys. {\bf A542}, 310
  (1992).

\bibitem{BF3} H. Esbensen, G. F.  Bertsch and K. Hencken,
  Phys. Rev. C{\bf 56}, 3054 (1999).

\bibitem{VMP96}N. Vinh Mau and J.C. Pacheco, Nucl. Phys. {\bf A607}, 
163 (1996).

\bibitem{BVM97}A. Bonaccorso and N. Vinh Mau, Nucl. Phys. {\bf A615}, 
245 (1997).

\bibitem{Zhukov93} M.V. Zhukov {\it et al.}, Phys. Rep. {\bf 231}, 151
(1993).

\bibitem{Fed95} D. V. Fedorov, E. Garrido and A. S. Jensen,
   Phys. Rev. C{\bf 51}, 3052 (1995).

\bibitem{Pud95} B. S. Pudliner, V. R. Pandharipande, J. Carlson and
  R. B. Wiringa, Phys. Rev. Lett. {\bf 74}, 4396 (1995); S. C. Pieper,
  V. R. Pandharipande, R. B. Wiringa and J. Carlson, Phys. Rev. C{\bf 64},
  014001 (2001).

\bibitem{Matsuo} M. Matsuo, K. Mizuyama and Y. Serizawa,
  Phys. Rev. C{\bf 71},  064326 (2005).

\bibitem{Aoyama} S. Aoyama, K. Kato and K. Ikeda, Prog. Theor. Phys.
  Suppl. {\bf 142}, 35 (2001); 
T. Myo, S. Aoyama, K. Kato and K. Ikeda, Prog. Theor. Phys. 
{\bf 108}, 133 (2002).

\bibitem{Myo}T. Myo, S. Aoyama, K. Kato, and K. Ikeda,
Phys. Lett. B{\bf 576}, 281 (2003); 
T. Myo, K. Kato, S. Aoyama, and K. Ikeda,
Phys. Rev. C{\bf 63}, 054313 (2001).

\bibitem{OZV99}Yu. Ts. Oganessian, V.I. Zagrebaev, and J.S. Vaagen, Phys. 
Rev. Lett. {\bf 82}, 4996 (1999); Phys. Rev. C{\bf 60}, 044605 (1999).

\bibitem{BDP00}K. Bennaceur, J. Dobaczewski, and M. Ploszajczak, 
Phys. Lett. {\bf B496}, 154 (2000).

\bibitem{STG92}H. Sagawa, N. Takigawa, and Nguyen Van Giai, 
Nucl. Phys. {\bf A543}, 575 (1992).

\bibitem{BBH91}C.A. Bertulani, G. Baur, and M.S. Hussein, 
Nucl. Phys. {\bf A526}, 751 (1991).

\bibitem{HHB03}K. Hagino, M.S. Hussein, and A.B. Balantekin, 
Phys. Rev. C{\bf 68}, 048801 (2003). 
\bibitem{Li11_exp}
  K. Ieki {\it et al.}, Phys. Rev. Lett. {\bf 70}, 730 (1993);
  S. Shimoura {\it et al.}, Phys. Lett. {\bf B348}, 29 (1995);
  M. Zinser {\it et al.}, Nucl. Phys. {\bf A619}, 151 (1997). 
\bibitem{NF05}T. Nakamura and N. Fukuda, Eur. Phys. J. A, in 
press (2005).

\end{thebibliography}
\end{document}